\documentclass[aps,pra,twocolumn,showpacs,floatfix,longbibliography]{revtex4-2}
\usepackage{graphicx,amsfonts,amssymb,amsmath}
\usepackage{textcomp}
\usepackage{esint}
\usepackage[english]{babel}
\usepackage{afterpage}
\usepackage{bbold}
\usepackage{soul}
\usepackage[colorlinks,linktocpage,citecolor=blue,linkcolor=red,urlcolor=blue]{hyperref}

\input cyracc.def

\allowdisplaybreaks

\newcommand{\Shop}{S_{\rm hop}}
\newcommand{\SLL}{S_{\rm LL}}
\def\varphiset{\{\varphi_{a}\}}
\def\Jset{\{J_{a}\}}
\def\Jfset{\{J_{f}\}} 
\def\phiset{\{\phi_{a}\}}
\newcommand{\Kkpure}{K_k^{\rm (pure)}}

\newcommand{\KLL}{K_{\rm LL}}
\newcommand{\vLL}{v_{\rm LL}}

\begin{document}
	

\newcommand{\oldnew}[2]{\marginpar{\scriptsize \textcolor{red}{correction}}{\textcolor{red}{#2}}}
\newcommand{\suppressed}[1]{\marginpar{\scriptsize \textcolor{red}{correction}}{\textcolor{red}{\st{#1}}}}
\newcommand{\correction}[1]{\marginpar{\textcolor{red}{\scriptsize #1}}}
\newcommand{\marge}[1]{\marginpar{\scriptsize #1}}
\newcommand{\remarque}[1]{\marginpar{\scriptsize Remarque}{\it [#1]}}

%

\def\rhoeq{\hat\rho_{\rm eq}}

\newcommand{\new}[1]{{\bf #1}}
\newlength{\textlarg}
\newcommand{\redbar}[1]{\textcolor{red}{\st{#1}}} 
\newcommand{\bluebar}[1]{\textcolor{blue}{\st{#1}}} 

\newcommand{\normord}[1]{:\mathrel{#1}:}

\newcommand{\beq}{\begin{equation}}
\newcommand{\eeq}{\end{equation}}
\newcommand{\bfig}{\begin{figure}}
\newcommand{\efig}{\end{figure}}
\newcommand{\bline}{\begin{multline}}
\newcommand{\eline}{\end{multline}}
\newcommand{\bremark}{\begin{quotation} \noindent \small }
\newcommand{\eremark}{\end{quotation}}
\newcommand{\llbrace}{\left\lbrace}  
\newcommand{\rrbrace}{\right\rbrace}
\newcommand{\lbraket}{\left[}
\newcommand{\rbraket}{\right]}
\newcommand{\llangle}{\left\langle}
\newcommand{\rrangle}{\right\rangle} 

\newcommand{\Tr}{{\rm Tr}} 
\newcommand{\tr}{{\rm tr}} 
\newcommand{\sgn}{\,{\rm sgn}} 
\newcommand{\mean}[1]{\langle #1 \rangle}
\newcommand{\commu}[2]{[#1,#2]} 
\newcommand{\bra}[1]{\langle#1|}
\newcommand{\ket}[1]{|#1\rangle}
\newcommand{\braket}[2]{\langle #1|#2\rangle}
\newcommand{\ketbra}[2]{|#1\rangle\langle#2|}
\newcommand{\dbraket}[3]{\langle #1|#2|#3\rangle}
\newcommand{\tens}[1]{\overleftrightarrow{#1}}  
\newcommand{\vac}{|{\rm vac}\rangle} 
\newcommand{\bravac}{\langle{\rm vac}|}
\newcommand{\const}{{\rm const}} 
\newcommand{\unif}{{\rm unif.}} 
\newcommand{\atanh}{\,{\rm atanh}}
\newcommand{\cotanh}{\,{\rm cotanh}}

\newcommand{\ie}{i.e.\xspace}
\newcommand{\iet}{i.e.}
\newcommand{\eg}{e.g.\xspace}
\newcommand{\cc}{{\rm c.c.}} 
\newcommand{\hc}{{\rm h.c.}} 
\newcommand{\etal}{{\it et al. }}
\newcommand\eme{$^{\mbox{\footnotesize ème}}$\xspace}

\newcommand{\jhatbf}{\hat {\textbf \jold}} 
\newcommand{\Jhatbf}{\hat {\textbf \J}} 
\newcommand{\jhat}{\hat {\jmath}} 
\newcommand{\Jhat}{\hat {J}} 
\newcommand{\jbf}{\textbf j}
\newcommand{\Jbf}{\textbf J}

\def\chibf{\boldsymbol{\chi}}
\def\down{\downarrow}
\def\eps{\epsilon}
\def\gam{\gamma} 
\def\alphabf{\boldsymbol{\alpha}}
\def\phibf{\boldsymbol{\phi}}
\def\varphibf{\boldsymbol{\varphi}}
\def\varphibfs{\boldsymbol{\varphi}_<}
\def\varphibfl{\boldsymbol{\varphi}_>}
\def\varphis{\varphi_{<}}
\def\varphil{\varphi_{>}}
\def\psibf{\boldsymbol{\psi}}
\def\thetabf{\boldsymbol{\theta}}
\def\Ome{\Omega}
\def\omeD{{\omega_D}} 
\def\bfOme{\boldsymbol{\Omega}} 
\def\Omebf{\boldsymbol{\Omega}} 
\def\lamb{\lambda}
\def\Lamb{\Lambda}
\def\sig{\sigma}
\def\Sig{\Sigma}
\def\sigp{{\sigma'}} 
\def\bfsig{\boldsymbol{\sigma}} 
\def\sigbf{\boldsymbol{\sigma}} 
\def\bfSig{\boldsymbol{\Sigma}} 
\def\The{\Theta} 
\def\up{\uparrow}

\def\epsk{\epsilon_{\bf k}} 
\def\epsp{\epsilon_{\bf p}} 
\def\xik{\xi_{\bf k}} 
\def\txik{\tilde\xi_{\bf k}} 
\def\xip{\xi_{\bf p}} 
\def\epsq{\epsilon_{\bf q}} 
\def\xiq{\xi_{\bf q}} 
\def\xikq{\xi_{{\bf k}+{\bf q}}} 
\def\Ek{E_{\bf k}} 
\def\Ep{E_{\bf p}}
\def\Eq{E_{\bf q}}
\def\Heff{\hat H_{\rm eff}}
\def\Hem{\hat H_{\rm em}}
\def\Hint{\hat H_{\rm int}}
\def\Hloc{\hat H_{\rm loc}}
\def\HMF{\hat H_{\rm MF}}
\def\HLL{\hat H_{\rm LL}}
\def\Hdis{\hat H_{\rm dis}}
\def\Sem{S_{\rm em}}
\def\SMF{S_{\rm MF}} 
\def\SHF{S_{\rm HF}} 
\def\SRPA{S_{\rm RPA}} 
\def\Sint{S_{\rm int}} 
\def\Sloc{S_{\rm loc}}
\def\TN{T_{\rm N}} 
\def\TNHF{T^{\rm HF}_{\rm N}} 
\def\Zloc{Z_{\rm loc}} 
\def\ZMF{Z_{\rm MF}} 
\def\ZHF{Z_{\rm HF}} 
\def\ZRPA{Z_{\rm RPA}} 
\def\RPA{{\rm RPA}}
\def\loc{{\rm loc}} 
\def\pp{{\rm pp}}
\def\ph{{\rm ph}} 
\def\ch{{\rm ch}}
\def\sp{{\rm sp}} 
\def\qtf{q_{\rm TF}}
\def\epstf{\eps^{}_{\rm TF}} 
\def\epsrpa{\eps^{}_{\rm RPA}} 
\def\chinnzpp{\chi_{nn}^{0}{}\!\!\!''}
\def\SigHF{\Sigma_{\rm HF}}
\def\psicl{\psi_{\rm cl}} 

\def\half{\frac{1}{2}}
\def\dhalf{\dfrac{1}{2}}
\def\third{\frac{1}{3}} 
\def\quarter{\frac{1}{4}}

\def\qr{{\bf q}\cdot{\bf r}}
\def\wt{\omega t} 

\def\a{{\bf a}}
\def\b{{\bf b}}
\newcommand{\cv}{{\bf c}} 
\def\e{{\bf e}}
\def\f{{\bf f}}
\def\g{{\bf g}}
\def\h{{\bf h}}
\def\jold{\char"11}
\def\j{{\bf j}}
\def\k{{\bf k}}
\def\l{{\bf l}}
\def\ellbf{\bm{\ell}} 
\def\m{{\bf m}}
\def\n{{\bf n}} 
\def\p{{\bf p}} 
\def\q{{\bf q}}
\def\r{{\bf r}}
\def\t{{\bf t}}
\def\u{{\bf u}}
\newcommand{\vv}{{\bf v}}
\def\x{{\bf x}}
\def\y{{\bf y}} 
\def\z{{\bf z}} 
\def\A{{\bf A}}
\def\B{{\bf B}}
\def\D{{\bf D}} 
\def\E{{\bf E}} 
\def\F{{\bf F}} 
\def\H{{\bf H}}  
\def\J{{\bf J}}
\def\K{{\bf K}} 

\def\G{{\bf G}}
\def\L{{\bf L}}
\def\M{{\bf M}}  
\def\O{{\bf O}} 
\def\P{{\bf P}} 
\def\Q{{\bf Q}} 
\def\R{{\bf R}}
\def\S{{\bf S}}
\def\U{{\bf U}} 
\def\V{{\bf V}} 
\def\X{{\bf X}} 
\def\Y{{\bf Y}} 
\def\epsbf{\boldsymbol{\epsilon}}
\def\betabf{\boldsymbol{\beta}}
\def\deltabf{\boldsymbol{\delta}}
\def\mubf{\boldsymbol{\mu}}
\def\nablabf{\boldsymbol{\nabla}}
\def\rhobf{\boldsymbol{\rho}}
\def\sigmabf{\boldsymbol{\sigma}} 
\def\Pibf{\boldsymbol{\Pi}}
\def\pibf{\boldsymbol{\pi}}

\def\para{\parallel}
\def\kpara{{k_\parallel}}
\def\kperp{{k_\perp}} 
\def\kperpp{{k_\perp'}} 
\def\qperp{{q_\perp}} 
\def\tperp{{t_\perp}} 

\def\w{\omega}
\def\wn{\omega_n}
\def\wm{\omega_m}
\def\wnu{\omega_\nu}
\def\wp{\omega_p} 
\def\dmu{{\partial_\mu}}
\def\dnu{{\partial_\nu}}
\def\dl{{\partial_l}}  
\def\dt{\partial_t} 
\def\tdt{\tilde\partial_t}
\def\dk{\partial_k}
\def\tdk{\tilde\partial_k}
\def\dx{\partial_x}
\def\dy{\partial_y} 
\def\dw{\partial_{\w}}
\def\dtau{{\partial_\tau}}  
\def\det{{\rm det}} 
\def\Pf{{\rm Pf}}
\def\diag{{\rm diag}}

\def\dsum{\displaystyle \sum}
\def\dint{\displaystyle \int} 
\def\intt{\int_{-\infty}^\infty dt} 
\def\inttp{\int_{-\infty}^\infty dt'} 
\def\intk{\int_{\bf k}} 
\def\intkd{\int \frac{d^dk}{(2\pi)^d}}
\def\intq{\int_{\bf q}} 
\def\intr{\int d^dr}  
\def\dintr{\displaystyle \int d^dr} 
\def\intrp{\int d^dr'}
\def\dinttau{\displaystyle \int_0^\beta d\tau}
\def\dinttaup{\displaystyle \int_0^\beta d\tau'}
\def\inttau{\int_0^\beta d\tau}
\def\inttaup{\int_0^\beta d\tau'}
\def\intx{\int d^{d+1}x} 
\def\inttaur{\int_0^\beta d\tau \int d^dr}
\def\intinf{\int_{-\infty}^\infty}
\def\dinttaur{\displaystyle \int_0^\beta d\tau \int d^dr}
\def\dintinf{\displaystyle \int_{-\infty}^\infty}
\def\intw{\int_{-\infty}^\infty \frac{d\w}{2\pi}}
\def\sumr{\sum_{\bf r}} 

\def\calA{{\cal A}}
\def\calAbf{\bm{{\cal A}}}
\def\calB{{\cal B}} 
\def\calC{{\cal C}} 
\def\dt{\partial_t}
\def\calD{{\cal D}}
\def\calE{{\cal E}}
\def\calF{{\cal F}} 
\def\calFbf{\bm{{\cal F}}}
\def\calG{{\cal G}}
\def\calH{{\cal H}}
\def\calI{{\cal I}}
\def\calJ{{\cal J}}
\def\calK{{\cal K}}
\def\calL{{\cal L}} 
\def\calM{{\cal M}} 
\def\calN{{\cal N}}
\def\calO{{\cal O}}
\def\calP{{\cal P}}  
\def\calR{{\cal R}} 
\def\calS{{\cal S}}
\def\calT{{\cal T}}
\def\calU{{\cal U}}
\def\calV{{\cal V}}
\def\calX{{\cal X}} 
\def\calY{{\cal Y}} 
\def\calW{{\cal W}} 
\def\calZ{{\cal Z}}

\def\tT{{\tilde T}}
\def\talpha{{\tilde\alpha}}
\def\tbeta{{\tilde\beta}}
\def\tchi{{\tilde\chi}}
\def\tdelta{{\tilde\delta}}
\def\tDelta{{\tilde\Delta}}
\def\teta{{\tilde\eta}} 
\def\tlamb{{\tilde\lambda}}
\def\tmu{{\tilde\mu}}
\def\tphibf{{\tilde\phibf}}
\def\trho{{\tilde\rho}}
\def\tvarphibf{{\tilde\varphibf}} 
\def\tq{\tilde q}
\def\tw{{\tilde\omega}}
\def\twn{{\tilde\omega_n}}
\def\twnu{{\tilde\omega_\nu}}

\def\asinh{{\rm asinh}} 
\def\Tbkt{T_{\rm BKT}}

\graphicspath{{./figures_submit/}}
	
\title{Superfluid--Bose-glass transition in a system of disordered bosons \\
with long-range hopping in one dimension}

\author{Nicolas Dupuis}
\affiliation{Sorbonne Universit\'e, CNRS, Laboratoire de Physique Th\'eorique de la Mati\`ere Condens\'ee, LPTMC, F-75005 Paris, France}
	
\date{September 11, 2024} 
	
\begin{abstract}
We study the superfluid--Bose-glass transition in a one-dimensional lattice boson model with power-law decaying hopping amplitude $t_{i-j}\sim 1/|i-j|^\alpha$, using bosonization and the nonperturbative functional renormalization group (FRG). When $\alpha$ is smaller than a critical value $\alpha_c<3$, the U(1) symmetry is spontaneously broken, which leads to a density mode with nonlinear dispersion and dynamical exponent $z=(\alpha-1)/2$; the superfluid phase is then stable for sufficiently weak disorder, contrary to the case of short-range hopping where disorder is a relevant perturbation when the Luttinger parameter is smaller than $3/2$. In the presence of disorder, however, long-range hopping has no effect in the infrared limit and the FRG flow eventually becomes similar to that of a boson system with short-range hopping. This implies that the superfluid phase, when stable, exhibits a density mode with linear dispersion ($z=1$) and the superfluid--Bose-glass transition remains in the Berezinskii-Kosterlitz-Thouless universality class, while the Bose-glass fixed point is insensitive to long-range hopping. We compare our findings with a recent numerical study.
\end{abstract}
	
\maketitle
 
\tableofcontents
	
\section{Introduction.} 

The ground state of a one-dimensional quantum fluid is generically a Luttinger liquid, i.e. a state characterized by a gapless spectrum and quasi-long-range order~\cite{Giamarchi_book}. When the particles interact {\it via} a long-range potential $V(x)\sim 1/|x|^\gamma$ with an exponent $\gamma<1$, the ground state is not a Luttinger liquid but a Wigner crystal with a spontaneous modulation of the density~\cite{Schulz93,Daviet20}. On the other hand, the ground state of the spin-$\half$ XXZ model with long-range hopping amplitude $t_{i-j}\sim 1/|i-j|^\alpha$ spontaneously breaks the U(1) symmetry when the exponent $\alpha$ is smaller than a critical value $\alpha_c<3$~\cite{Laflorencie05,Gong16,Maghrebi17,Frerot17}. 

In a Luttinger liquid, disorder has noticeable effects since it can induce a transition to a localized phase~\cite{Giamarchi87,Giamarchi88}, e.g. a Bose glass in the case of a Bose gas. What is the fate of the Luttinger liquid when both disorder and long-range interactions or hopping are present? In Ref.~\cite{Daviet20}, it was shown that the ground state of a Bose gas with long-range interactions is either a Wigner crystal or a Mott glass ---a phase intermediate between a Mott insulator and a Bose glass---, depending on the value of the exponent $\gamma<1$ characterizing the power-law decay of the interaction potential. In the case of long-range hopping, it has been recently suggested using quantum Monte Carlo methods that a one-dimensional boson system is always superfluid (at zero temperature) for a sufficiently weak disorder~\cite{Gupta23}, in contrast to the usual case of short-range hopping where disorder is a relevant perturbation when the Luttinger parameter $K$ is smaller than $3/2$~\cite{Giamarchi87,Giamarchi88}. Furthermore, the superfluid--Bose-glass transition would not be in the Berezinskii-Kosterlitz-Thouless (BKT) universality class ---as in the case of short-range hopping~\cite{Giamarchi87,Giamarchi88}--- but would behave like a usual continuous phase transition with a power-law divergence of the correlation length~\cite{Gupta23}.

In this paper, we study a one-dimensional disordered lattice boson model with hopping amplitude $t_{i-j}$ decaying with distance as $1/|i-j|^\alpha$. In Sec.~\ref{sec_lrh}, using bosonization we show that the U(1) symmetry is spontaneously broken when the exponent $\alpha$ is smaller than a critical value $\alpha_c<3$, in agreement with previous results obtained for the equivalent spin-$\half$ XXZ model with long-range hopping~\cite{Laflorencie05,Gong16,Maghrebi17,Frerot17}. It is possible to describe this system in the usual Luttinger-liquid formalism provided that we consider a Luttinger parameter $K(q)$ and a velocity $v(q)$ dependent on momentum. The small-$q$ divergence $K(q),v(q)\sim |q|^{(\alpha-3)/2}$ is responsible for the spontaneous breaking of the U(1) symmetry. It also leads to a non-linear dispersion $\w\sim |q|^{(\alpha-1)/2}$, with a dynamical exponent $z=(\alpha-1)/2$, of the density mode and to a diverging momentum-dependent superfluid stiffness $\rho_s(q)\sim |q|^{\alpha-3}$ in the limit $q\to 0$. In Sec.~\ref{sec_dis}, we study the effect of disorder using the replica formalism and the nonperturbative functional renormalization group (FRG). Contrary to the short-range hopping case, where the superfluid phase is unstable for any nonzero disorder when the Luttinger parameter $\KLL$ is smaller than $3/2$, we find that the superfluid phase is always stable for sufficiently weak disorder. This can be understood by the low-momentum divergence of $K(q)$, which implies that the criterion $K(q)>3/2$ will always be fulfilled for a sufficient small value of $q$. However, the long-range hopping is not resistant to disorder and in the low-energy limit one always recovers the standard FRG flow equations of a disordered boson system with short-range hopping, i.e. of a disordered Luttinger liquid with a well-defined zero-momentum limit $K(q\to 0)$ of the Luttinger parameter. This implies that the superfluid phase, when stable, exhibits a density mode with a linear dispersion $\w\sim|q|$, corresponding to a dynamical exponent $z=1$, and that the superfluid--Bose-glass transition is in the BKT universality class. The Bose-glass fixed point~\cite{Dupuis19,Dupuis20} is also insensitive to long-range hopping. We discuss the difference between our findings and those of Ref.~\cite{Gupta23} in the conclusion.

\section{Boson lattice model with long-range hopping} 
\label{sec_lrh}

We consider a lattice boson model with Hamiltonian 
\beq 
\hat H = - \sum_{i<j} t_{i,j} ( \hat\psi^\dagger_i \hat\psi_j + \hc ) + \frac{U}{2} \sum_i \hat\psi^\dagger_i  \hat\psi^\dagger_i \hat\psi_i  \hat\psi_i  , 
\label{ham}
\eeq
where the intersite hopping amplitude $t_{i,j}$ between sites $i$ and $j$ decays as a power law, 
\beq
t_{i,j} = \frac{t}{|i-j|^\alpha} , 
\eeq
with an exponent $\alpha>1$. This condition ensures energy extensivity in the thermodynamic limit. In Eq.~(\ref{ham}), $\hat\psi^\dagger_i$ and $\hat\psi_i$ denote boson creation and annihilation operators at site $i$. The Euclidean (imaginary-time) action reads~\cite{NDbook1} 
\beq 
S = \inttau \biggl\{ \sum_i \psi_i(\tau)\dtau\psi_i(\tau) + H[\psi^*(\tau),\psi(\tau)] \biggr\} , 
\eeq 
where $\psi_i(\tau)=\psi_i(\tau+\beta)$ is a complex field satisfying periodic boundary conditions in time. Throughout the paper we set $\hbar$, $k_B$ and the lattice spacing to unity and consider only the zero-temperature limit $\beta=1/T\to\infty$.

\subsection{Bosonization} 

We use a density-phase representation where the boson field 
\beq 
\psi_j(\tau) = \sqrt{\rho_j(\tau)} e^{i\theta_j(\tau)}
\eeq 
is written in terms of the density $\rho_j=|\psi_j|^2$ (i.e. the number of bosons at site $i$) and the phase $\theta_j $. This leads to the action $S=\Sloc+\Shop$, where the local part is given by
\beq 
\Sloc = \inttau \biggl( \sum_j i\rho_j \dtau\theta_j + \frac{U}{2} \rho_j^2  \biggr) 
\eeq 
and the hopping part by
\beq  
\Shop = - \sum_{i\neq j} t_{i,j} \inttau  \sqrt{\rho_i\rho_j} \cos(\theta_i-\theta_j) .
\eeq 
Expanding $\rho_i=\rho_0+\delta\rho_i$ about the mean density $\rho_0$, we rewrite the hopping part of the action as 
\begin{align}
\Shop ={}& - \inttau  \sum_{i\neq j} t_{i,j} \Bigl[  \rho_0 + \half (\delta\rho_i + \delta\rho_j ) \nonumber\\ & - \frac{1}{8\rho_0} (\delta\rho_i - \delta\rho_j )^2\Bigr] \cos(\theta_i-\theta_j)  
\end{align}
to second order in $\delta\rho_i$. 

We now consider the continuum limit and follow the standard bosonization rules where the long-wavelength part of the density fluctuation $\delta\rho_i\to \delta\rho(x)=- \frac{1}{\pi}\nabla\varphi(x)$ is written in terms of a phase field $\varphi$~\cite{,Haldane81,Giamarchi_book}. Retaining only the leading contribution in derivatives, we obtain the action
\beq 
\Shop = - g \int_{x,y,\tau} \frac{\cos(\theta(x,\tau)-\theta(y,\tau))}{|x-y|^\alpha} , 
\label{Shop} 
\eeq 
where $g=t\rho_0$, and we use the notation $\int_\tau\equiv\inttau$ and $\int_x\equiv \int dx$. A short-distance cutoff, of the order of the lattice spacing, is implied. The local part of the action takes the form 
\beq 
\Sloc = \int_{x,\tau} \biggl[ \frac{U}{2\pi^2} (\nabla\varphi)^2 - \frac{i}{\pi} \nabla\varphi \dtau \theta \biggr] .
\eeq 

The procedure we have followed so far is very crude but allows one to determine the low-energy form of the action $\Shop$ associated with long-range hopping. The short-distance physics, which is not included in this naive bosonization, can be taken into account by writing the full action as $\Shop+\SLL$, where $\Shop$ is given by~(\ref{Shop}) and $\SLL$ is the usual action of a Luttinger liquid with Luttinger parameter $\KLL\equiv \KLL(\alpha)$ and velocity $\vLL\equiv \vLL(\alpha)$, i.e. 
\begin{align}
S ={}& \int_{x,\tau} \biggl\{ \frac{\vLL}{2\pi} \left[ \frac{1}{\KLL} (\nabla\varphi)^2 + \KLL (\nabla\theta)^2 \right]  - \frac{i}{\pi} \nabla\varphi \dtau \theta \biggr\} \nonumber\\ 
& - g \int_{x,y,\tau} \, \frac{\cos(\theta(x,\tau)-\theta(y,\tau))}{|x-y|^\alpha} .
\label{action1} 
\end{align} 
This action must be supplemented with a UV momentum cutoff $\Lambda$ of the order of the inverse lattice spacing. We can check the relevance or irrelevance of $g$ at the Luttinger-liquid fixed point (described by the action $\SLL$) by dimensional analysis, 
\beq 
[g] = 3 - \alpha - 2 [e^{i\theta}]_{\rm LL} = 3 - \alpha - \frac{1}{2\KLL} , 
\eeq 
where the last result follows from $[e^{i\theta}]_{\rm LL}=1/4\KLL$. The cosine term in~(\ref{action1}) is therefore relevant if
\beq 
\alpha < 3 - \frac{1}{2\KLL} ,
\label{relevance}
\eeq 
where we stress that $\KLL\equiv \KLL(\alpha)$ is a function of $\alpha$ if the low-energy effective action~(\ref{action1}) is derived from the lattice boson model~(\ref{ham}). In the following, we consider $\KLL$ and $\alpha$ as two independent parameters. 

When the condition~(\ref{relevance}) is not fulfilled, the long-range hopping is irrelevant and the system behaves as a (standard) Luttinger liquid in the low-energy limit. In the opposite case, $g$ is a relevant perturbation and one can expand the cosine term to quadratic order, which gives the action 
\beq 
S = \SLL+ \frac{g}{2} \int_{x,y,\tau} \frac{(\theta(x,\tau)-\theta(y,\tau))^2}{|x-y|^\alpha} .
\label{action2} 
\eeq
In Fourier space, the non-local term in~(\ref{action2}) is written as 
\beq 
\frac{g}{2} \sum_{Q} ( c_2 q^2 +c_{\alpha-1}|q|^{\alpha-1} )|\theta(Q)|^2 
\eeq 
to quadratic order in $q$, where $Q=(q,i\wn)$ and $\wn=2n\pi T$ ($n$ integer) is a bosonic Matsubara frequency. In the zero-temperature limit, $\wn$ becomes a continuous variable that will be simply denoted by $\w$. When $1<\alpha<3$, the coefficient $c_2$ is negative and $c_{1-\alpha}$ positive; their exact expression is not important for our discussion. The quadratic term $c_2q^2$ can be taken into account by a redefinition of $\vLL$ and $\KLL$: The ratio $\vLL/\KLL$ is unchanged but $\vLL\KLL$ is reduced by $\pi g|c_2|$. This leads to 
\begin{align}
S ={}& \sum_{Q} \biggl\{ \frac{\vLL}{2\pi} q^2 \biggl[ \frac{1}{\KLL} |\varphi(Q)|^2 + \KLL |\theta(Q)|^2 \biggr] \nonumber\\ &
+ \frac{i}{\pi} q\w \varphi(-Q) \theta(Q) 
+ \half Z_s |q|^{\alpha-1} |\theta(Q)|^2 \biggr\} ,
\end{align}  
where $Z_s=g c_{\alpha-1}$. This action can be put in the canonical Luttinger-liquid form,
\begin{align}
S ={}& \sum_{Q} \biggl\{ \frac{v(q)}{2\pi} q^2 \biggl[ \frac{1}{K(q)} |\varphi(Q)|^2 + K(q) |\theta(Q)|^2 \biggr] \nonumber\\ & + \frac{i}{\pi} q\w \varphi(-Q) \theta(Q) \biggr\} ,
\label{action3} 
\end{align}
if one introduces a velocity and a Luttinger parameter dependent on $q$,
\beq
\begin{split}
\frac{v(q)}{K(q)} &= \frac{\vLL}{\KLL} , \\ 
v(q) K(q) &= \vLL\KLL + \pi Z_s |q|^{\alpha-3} . 
\end{split}
\label{Kvqdef} 
\eeq 
The action being quadratic in $\theta$, one can integrate out the latter to obtain the $\varphi$-only action 
\beq 
S = \sum_Q |\varphi(Q)|^2 \frac{v(q)}{2\pi K(q)} \left( q^2 + \frac{\w^2}{v(q)^2} \right) ,
\label{action4} 
\eeq 
a form that will be useful when studying the system in the presence of disorder. 

From Eqs.~(\ref{Kvqdef}), we deduce 
\beq
K(q) = \KLL \left( 1 +  \frac{\pi Z_s}{\KLL\vLL}  |q|^{\alpha-3} \right)^{1/2} . 
\label{Kqdef}
\eeq 
This expression defines a crossover scale 
\beq 
q_x = \left( \frac{\pi Z_s}{\KLL\vLL} \right)^{1/(3-\alpha)}
\eeq 
separating a high-momentum regime where $K(q)\simeq \KLL$ from a low-momentum regime dominated by long-range hopping, $K(q)\sim |q|^{(\alpha-3)/2}$. This crossover scale is meaningful only if $q_x\ll\Lambda$. It is difficult to estimate its value in the lattice model defined by~(\ref{ham}). On the one hand, one must estimate $\KLL$ and $\vLL$ (which are functions of $\alpha$), and then one has to take into account the (negative) coefficient $c_2$ which reduces the value of $\vLL\KLL$ (see the preceding discussion). In the following we therefore consider $\KLL$, $\vLL$ and $Z_s$, as well as the UV momentum cutoff $\Lambda$, as independent parameters. Note that this does affect the nature of the superfluid--Bose-glass transition but may change the details of the phase diagram as discussed in Sec.~\ref{subsubsection_flow_diagram}. 

The actions~(\ref{action3}) and (\ref{action4}), with the definition~(\ref{Kvqdef}) of the $q$-dependent velocity and Luttinger parameter, describe the low-energy limit of a boson system with long-range hopping. Since $v(q),K(q)\sim |q|^{(\alpha-3)/2}$ diverges when $q\to 0$, one expects the low-energy properties to be markedly different from those of a boson system with short-range hopping behaving as a  Luttinger liquid in the low-energy limit.

\subsection{Low-energy properties}

The actions~(\ref{action3}) and (\ref{action4}) being quadratic, they can be straightforwardly diagonalized. One finds a mode with dispersion $\w=v(q)|q|$ behaving as $|q|^{(\alpha-1)/2}$ in the long-wavelength limit, which gives a dynamical exponent $z=(\alpha-1)/2$. The compressibility is not affected by long-range hopping, 
\beq 
\kappa = \lim_{q\to 0} \frac{K(q)}{\pi v(q)} = \frac{\KLL}{\pi \vLL} , 
\eeq 
whereas the $q$-dependent superfluid stiffness 
\beq 
\rho_s(q) = \frac{v(q)K(q)}{\pi} \sim |q|^{\alpha-3} \quad (q\to 0) 
\eeq 
diverges in the small-momentum limit. This latter property can be seen as an enhancement of the superfluid character of the bosons due to long-range hopping. The conductivity  
\begin{equation} 
	\Re[\sig(q,\w)] 
	= \frac{D(q)}{2} [ \delta(\w-v(q)q) +\delta(\w+v(q)q) ] 
\label{Resig} 
\end{equation}
exhibits two Dirac peaks at $\w=\pm v(q)q$, as in a standard Luttinger liquid, but the Drude weight $D(q)=v(q)K(q)=\pi \rho_s(q)$ diverges when $q\to 0$ (see Appendix~\ref{app} for details). 

The expectation value $\mean{\psi(x,\tau)}\simeq \rho_0 \mean{e^{i\theta(x,\tau)}}$ of the boson field does not vanish showing that the U(1) symmetry of the boson system is spontaneously broken, 
\beq 
\mean{e^{i\theta(x,\tau)}} = e^{-\half \mean{\theta(x,\tau)^2}} > 0 . 
\label{U1broken} 
\eeq
The (connected) superfluid correlation function
\begin{align}
G(x) &= \mean{e^{i\theta(x,\tau)-i\theta(0,\tau)}}_c 
\sim 1/|x|^{\frac{3-\alpha}{2}} 
\label{GSFc} 
\end{align}
decays as a power law with an exponent $(3-\alpha)/2$. On the other hand, the density-density correlation function with momentum $q\simeq 2\pi\rho_0$, 
\beq 
\chi(x)=\mean{e^{2i\varphi(x,\tau)-2i\varphi(0,\tau)}} \sim e^{- \const\times |x|^{(3-\alpha)/2}} , 
\label{chi} 
\eeq 
decays as a stretched exponential, to be compared with the power-law decay $\sim 1/|x|^{2\KLL}$ in a Luttinger liquid. While in a standard superfluid density fluctuations at wavevectors $q\simeq \pm 2\pi\rho_0$ dominates over superfluid fluctuations when $\KLL<1/2$, the later are always the dominant fluctuations in the presence of long-range hopping. We conclude that long-range hopping enhances the superfluid character of the boson system, leading in particular to a spontaneous breaking of the U(1) symmetry, and tends to suppress density fluctuations at wave vector $q\simeq\pm 2\pi\rho_0$. The spontaneous symmetry breaking and the expression of the correlation functions are in agreement with the results of Refs.~\cite{Maghrebi17,Frerot17}, obtained from various analytical and numerical methods, once translated in the language of the equivalent spin-$\half$ models studied in these works.

\section{Long-range hopping and disorder}
\label{sec_dis}

We now consider a random potential (i.e. a random on-site energy in the lattice model~(\ref{ham})) which couples to the density of bosons. In the continuum limit, the disorder contributes to the action a term~\cite{Giamarchi87,Giamarchi88} 
\beq 
S_{\rm dis} = \int_{x,\tau} \left[ - \frac{1}{\pi} \nabla\eta + \rho_2 (\xi^* e^{2i\varphi} + \cc ) \right] , 
\eeq 
where $\eta(x)$ (real) and $\xi(x)$ (complex) denote random potentials with Fourier components near 0 and $\pm 2\pi\rho_0$, respectively. $\rho_2$ is a nonuniversal parameter that depends on microscopic details. $\eta(x)$ can be eliminated by a shift of $\varphi$ and is not considered in the following. The random potential $\xi$ is assumed to be Gaussian correlated with zero mean and variance 
\beq 
\overline{\xi^*(x)\xi(x')} = D \delta(x-x') 
\label{xivar} 
\eeq 
(all over averages vanish), where the overbar denotes the average over disorder. The full action of the system is thus given by $S+S_{\rm dis}$ where $S$ is the action in the absence of disorder [Eq.~(\ref{action4})]. In the replica formalism, one introduces $n$ copies of the system~\cite{Giamarchi88,Tarjus19,Dupuis20}. Averaging over disorder using~(\ref{xivar}) and integrating out the $\theta$ field then yields the replicated action 
\begin{multline} 
S[\{\varphi_a\}] = \sum_{a,Q} |\varphi_a(Q)|^2 \frac{v(q)}{2\pi K(q)} \left( q^2 + \frac{\w^2}{v(q)^2} \right) \\ 
 - \calD \sum_{a,b} \int_{x,\tau,\tau'} \cos[2\varphi_a(x,\tau)-2\varphi_b(x,\tau')] ,
\label{action5}
\end{multline}
where $\calD=D\rho_2^2$ and $a,b=1,\dots,n$ are replica indices.

\subsection{FRG formalism} 

To implement the nonperturbative FRG approach~\cite{Berges02,Kopietz_book,Delamotte12,Dupuis_review}, we add to the action~(\ref{action5}) the infrared regulator term~\cite{Dupuis20}
\begin{equation}
	\Delta S_k[\varphiset] = \half \sum_{a,Q} \varphi_{a}(-Q) R_{k}(Q) \varphi_{a}(Q) ,
	\label{DeltaSk} 
\end{equation}
where $k$ is a (running) momentum scale varying from the UV scale $\Lambda$ down to zero. The cutoff function $R_k(Q)$ is chosen so that fluctuation modes satisfying $|q|,|\w|/v_k\ll k$ are suppressed while those with $|q|\gg k$ or $|\w|/v_k\gg k$ are left unaffected (the $k$-dependent sound-mode velocity $v_k$ is defined below). Its precise form will be given in the following. 

The partition function
\begin{align}
	\calZ_k[\Jset] ={}& \int \calD[\varphiset] \exp \Bigl\{ -S[\varphiset] \nonumber \\ & 
	- \Delta S_k[\varphiset]+ \sum_{a} \int_{x,\tau} J_{a}\varphi_{a} \Bigr\} 
\end{align}
thus becomes $k$ dependent. The expectation value of the field, 
\beq
\phi_{a}(x,\tau) = \frac{\delta\ln \calZ_k[\Jfset]}{\delta J_{a}(x,\tau)} =\mean{\varphi_{a}(x,\tau)} ,
\eeq 
is obtained from a functional derivative with respect to the external source $J_a$ (to avoid confusion in the indices we denote by $\Jfset$ the $n$ external sources). 
The scale-dependent effective action 
\beq
\Gamma_k[\phiset] = - \ln \calZ_k[\Jset] + \sum_{a} \int_{x,\tau} J_{a} \phi_{a} - \Delta S_k[\phiset]
\eeq
is defined as a modified Legendre transform which includes the subtraction of $\Delta S_k[\phiset]$. Assuming that for $k=\Lamb$ the fluctuations are completely frozen by the term $\Delta S_{\Lamb}$, $\Gamma_{\Lamb}[\phiset]=S[\phiset]$. On the other hand the effective action of the original model~(\ref{action5}) is given by $\Gamma_{k=0}$ provided that $R_{k=0}$ vanishes. The nonperturbative FRG approach aims at determining $\Gamma_{k=0}$ from $\Gamma_{\Lamb}$ using Wetterich's equation~\cite{Wetterich93,Ellwanger94,Morris94} 
\beq
\dt \Gamma_k[\phiset] = \half \Tr \left\{ \dt R_k \bigl(\Gamma_k^{(2)}[\phiset] + R_k \bigr)^{-1} \right\} ,
\label{eqwet}
\eeq
where $\Gamma_k^{(2)}$ is the second functional derivative of $\Gamma_k$ and $t=\ln(k/\Lamb)$ a (negative) RG ``time''. The trace in~(\ref{eqwet}) involves a sum over momenta and frequencies as well as replica indices. 

To solve (approximately) the flow equation~(\ref{eqwet}) we consider the following ansatz for the effective action~\cite{Dupuis19,Dupuis20},
\beq 
\Gamma_k[\phiset] = \sum_a \Gamma_{1,k}[\phi_{a}] - \half \sum_{a,b}\Gamma_{2,k}[\phi_{a},\phi_{b}] ,
\label{ansatz1}
\eeq 
where 
\begin{align} 
	&\Gamma_{1,k}[\phi_{a}] = \half \sum_Q |\phi_a(Q)|^2 \bigl[ Z_x q^2 + Z_{\tau,k}(q) \w^2 ]  , \nonumber \\ 
	&\Gamma_{2,k}[\phi_a,\phi_b] = \int_{x,\tau,\tau'} V_{k} \bigl(\phi_{a}(x,\tau) - \phi_{b}(x,\tau') \bigr) ,
	\label{ansatz2}
\end{align} 
with initial conditions 
\beq 
Z_x=\frac{\vLL}{\pi \KLL}, \quad Z_{\tau,\Lambda}(q)=\frac{1}{\pi v(q) K(q)}
\eeq 
and
\beq 
V_{\Lamb}(u)=2\calD\cos(2u) .
\eeq 
The statistical tilt symmetry, due to the invariance of the disorder part of the action~(\ref{action5}) in the time-independent shift $\varphi_a(x,\tau)\to\varphi_a(x,\tau)+w(x)$ with $w(x)$ an arbitrary function of $x$, implies that $Z_x$ is not renormalized~\cite{Dupuis20}. On the other hand, we shall see that $Z_{\tau,\Lambda}(q)$ is renormalized only by an additive ($q$-independent) constant, 
\beq 
Z_{\tau,k}(q) = Z_{\tau,\Lambda}(q) + Y_k = \frac{1}{\pi v(q) K(q)} + Y_k . 
\label{Ykdef} 
\eeq 
It is convenient to write the one-replica effective action as 
\beq
\Gamma_1[\phi_a] = \sum_Q |\phi_a(Q)|^2 \frac{v_k(q)}{2\pi K_k(q)} \left( q^2 + \frac{\w^2}{v_k(q)^2} \right) , 
\eeq 
where 
\beq 
\begin{split}
\frac{v_k(q)}{K_k(q)} &= \frac{v(q)}{K(q)} = \frac{\vLL}{\KLL} , \\ 
\frac{1}{v_k(q)K_k(q)} &=\frac{1}{v(q)K(q)} + \pi Y_k . 
\end{split}
\eeq 
We further define 
\beq 
v_k = v_k(q=k), \quad K_k = K_k(q=k) , 
\eeq 
which represent the characteristic velocity and Luttinger parameter at the running momentum scale $k$. The velocity $v_k$ is used to define the cutoff function appearing in the infrared regulator term~(\ref{DeltaSk}),   
\beq 
R_k(Q) = Z_x \left( q^2 + \frac{\w^2}{v_k^2} \right) r\left( \frac{q^2}{k^2} + \frac{\w^2}{v_k^2 k^2}  \right) ,
\label{Rdef} 
\eeq 
where $r(y)=\gamma/(e^y-1)$ with $\gamma$ a constant of order unity.   

In the absence of disorder, the scale-dependent Luttinger parameter $\Kkpure=K(q=k)=\KLL v^{\rm (pure)}_k/\vLL$ is obtained from~(\ref{Kqdef}), i.e. 
\beq 
\Kkpure = \KLL \left( 1 +  \frac{\pi Z_s}{\KLL\vLL}  k^{\alpha-3} \right)^{1/2} .
\label{Kkpure} 
\eeq 
It reduces to $\KLL$ when $q_x\ll k\leq\Lambda$ and is fully determined by long-range hopping, i.e. $\Kkpure\sim k^{(\alpha-3)/2}$, when $k\ll q_x$.

\subsection{Flow equations}

The flow equations are best written in terms of the dimensionless two-replica potential~\cite{Dupuis20}
\beq 
\tilde V_k(\phi_a-\phi_b) = \frac{\KLL^2}{\vLL^2} \frac{V_k(\phi_a-\phi_b)}{k^3} . 
\eeq 
Furthermore we introduce the $k$-dependent dynamical exponent $z_k$ defined by 
\beq 
z_k - 1 = \dt \ln v_k , 
\label{zkdef} 
\eeq 
as well as the exponent $\theta_k$, 
\beq 
\theta_k = \dt \ln K_k = z_k - 1 , 
\eeq 
with the last result coming from $v_k/K_k=\vLL/\KLL$ and~(\ref{zkdef}). The derivation of the RG equation of the two-replica potential is standard~\cite{Dupuis20} and gives 
\begin{align}
\dt \delta_k(u) ={}& -3\delta_k(u) - K_k l_1 \delta''_k(u) \nonumber\\ & + \pi \bar l_2 \bigl\{ \delta''_k(u) [ \delta_k(u) - \delta_k(0) ]  + \delta_k'(u)^2 \bigr\} ,
\end{align} 
where $\delta_k(u)=-\tilde V_k''(u)$ and $l_1$ and $\bar l_2$ are  $k$-dependent threshold functions defined in Appendix~\ref{app_threshold}. The function $\delta_k(u)$ is $\pi$-periodic and can be expanded in circular harmonics, 
\beq 
\delta_{k}(u) =\sum_{n=1}^\infty \delta_{n,k} \cos(2nu) , 
\eeq 
a form which is useful for the numerical solution of the flow equations (with an upper cutoff $n_{\rm max}$ on the number of harmonics). 

As for the flow of $Z_{\tau,k}(q)$ [Eq.~(\ref{Ykdef})], we find 
\beq 
\dt Y_k = - \frac{1}{v_kK_k} \delta_k''(0) \bar m_\tau , 
\eeq 
where $\bar m_\tau$ is another ($k$-dependent) threshold function, which leads to 
\begin{align}
\theta_k ={}& \frac{\alpha-3}{2} (1-\pi v_k K_k Y_k) \left[ 1 - \frac{\KLL^2}{K_k^2} (1-\pi v_k K_k Y_k) \right] \nonumber\\ 
& + \frac{\pi}{2} \delta_k''(0) \bar m_\tau .  
\label{thetak} 
\end{align} 
In the absence of disorder, $Y_k=0$ and $K_k=\Kkpure$, one obtains $\theta_k=(\alpha-3)/2$ for $k\to 0$ in agreement with $\Kkpure\sim k^{(\alpha-3)/2}$. 

\begin{figure*}
	\includegraphics[width=4.12cm]{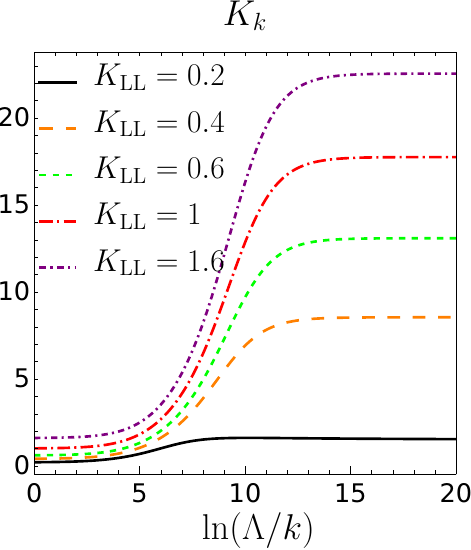}
	\includegraphics[width=4.2cm]{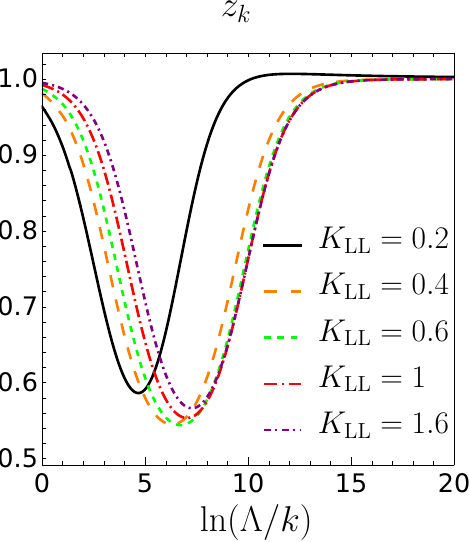}
    \includegraphics[width=4.15cm]{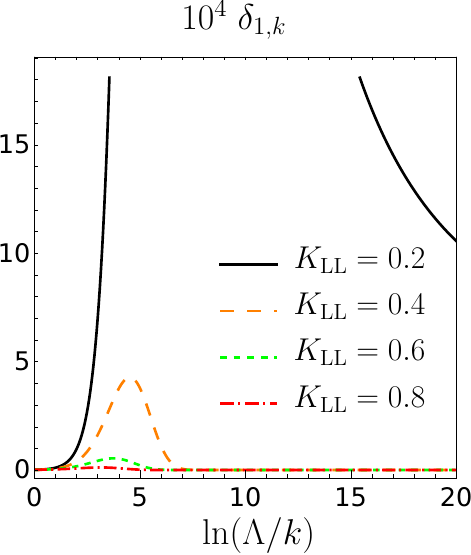}
	\includegraphics[width=4.2cm]{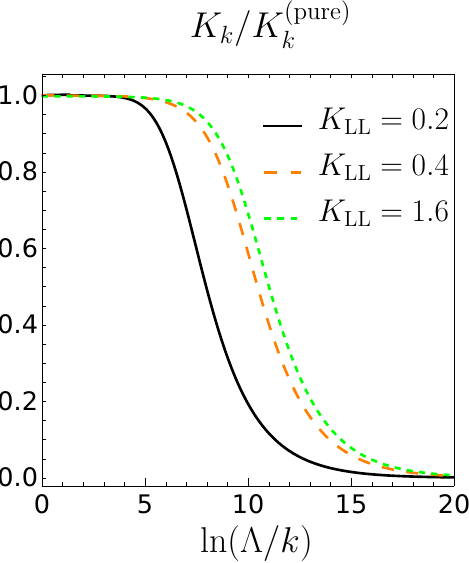}
	\caption{Luttinger parameter $K_k$, dynamical exponent $z_k$, amplitude $\delta_{1,k}$ of the first harmonic of $\delta_k(u)$, and ratio $K_k/\Kkpure$ vs $\ln(\Lambda/k)$ for a very weak value of the disorder ---corresponding to $\delta_{1,\Lambda}=10^{-6}$--- and various values of $\KLL$ in the range 0.2--1.6 ($\alpha=2$).} 
	\label{fig_weakD} 
\end{figure*}

\subsection{Qualitative analysis} 

The stability of the superfluid phase in the limit of an infinitesimal disorder is usually studied by considering the scaling dimension of the disorder variance $\calD$ in~(\ref{action5}). In a Luttinger liquid with dynamical exponent $z=1$ and Luttinger parameter $\KLL$,  $[e^{2i\varphi}]=\KLL$, so that $[\calD]=1+2z-2[e^{2i\varphi}]=3-2\KLL$, which indicates that the superfluid phase is stable for $\KLL>3/2$ but unstable when $\KLL<3/2$~\cite{Giamarchi87,Giamarchi88,Dupuis20}. A superfluid with long-range hopping is described by a scale-dependent Luttinger parameter $K(q)\sim |q|^{(\alpha-3)/2}$ ---or, equivalently, $\Kkpure\sim k^{(\alpha-3)/2}$--- which diverges in the long-wavelength limit. The (scale-dependent) scaling dimension $[\calD]=3-2\Kkpure$ will therefore always become negative in the small-$k$ limit thus making the disorder irrelevant. We conclude that, for sufficiently weak disorder, the superfluid phase is stable.

How is this conclusion modified when the disorder strength has an arbitrary value? In that case it is not possible to ignore the term $Y_k$ in $Z_{\tau,k}$ [Eq.~(\ref{Ykdef})], and the $q$-dependent velocity $v_k(q)$ at scale $k$ is given by 
\beq 
\frac{1}{v_k(q)^2} = \frac{1}{v(q)^2} + \frac{\pi \KLL}{\vLL} Y_k . 
\eeq 
Since $v(q)\sim |q|^{(\alpha-3)/2}$ diverges, the velocity $v_k(q)\to (\vLL/\pi \KLL Y_k)^{1/2}$ takes a finite value for $q\to 0$, which is independent of the long-range hopping encoded in $v(q)$ and $K(q)$. In other words, long-range hopping (and the associated enhanced superfluid character) is not resistant to disorder, and at sufficiently small scale $k$, one recovers the flow of a standard disordered Luttinger liquid with a scale-dependent Luttinger parameter $K_k=(\KLL/\pi \vLL Y_k)^{1/2}$. This implies, as will be further discussed in the following section, that long-range hopping cannot modify the universality class of the superfluid--Bose-glass transition. 

To fully characterize the flow, it is not sufficient to consider the running Luttinger parameter $K_k$. One must also identify the crossover between the short-distance regime where disorder is not important ---the system then becomes dominated by long-range hopping when $k\lesssim q_x$--- and the long-distance regime where disorder effectively suppresses long-range hopping. This can be achieved by considering the ratio $K_k/\Kkpure$, where $\Kkpure$ is the ($k$-dependent) Luttinger parameter in the pure system [Eq.~(\ref{Kkpure})].

\subsection{Numerical solution of the flow equations}

\subsubsection{Stability of the superfluid phase}

In Fig.~\ref{fig_weakD}, we show the flow of the Luttinger parameter $K_k$, the dynamical exponent $z_k$, the amplitude $\delta_{1,k}$ of the first harmonic of $\delta_k(u)$, and the ratio $K_k/\Kkpure$ obtained for $\alpha=2$, a very weak value of the disorder and various values of $\KLL$ in the range 0.2--1.6. All figures are obtained for $\Lambda\simeq 20$, $\vLL=1$ and $Z_s=0.1$.
These (arbitrary) values ensure that the momentum crossover scale $q_x$ is much smaller than $\Lambda$ so that the early stages of the flow is always dominated by short-range hopping. 

The initial conditions of the flow correspond to $K_\Lambda=K^{\rm (pure)}_\Lambda\simeq \KLL$ and $z_\Lambda\simeq 1$. In the beginning of the flow, the disorder plays essentially no role and $K_k\simeq \Kkpure$ increases as $k$ decreases. Disorder then starts to play an important role; the amplitude $\delta_{1,k}$ increases, $z_k$ decreases and $K_k$ begins to deviate from $\Kkpure$. In all cases however, the Luttinger parameter ends up being larger than $3/2$ and $\delta_{1,k}$ decreases towards zero. Thus, for the chosen value of the disorder, the system is in the superfluid phase for all values $0.2\leq \KLL\leq 1.6$. Regardless of the value of $\KLL$, it is always possible to choose a sufficiently small value of the disorder strength for the system to be in the superfluid phase. However, we note that in the superfluid phase $z_k\to 1$ when $k\to 0$ for any nonzero value of the disorder; the low-energy mode with dispersion $\w\sim |q|^{(\alpha-1)/2}$ of the pure system is suppressed and replaced by the usual sound mode with linear dispersion $\w\sim |q|$.

\subsubsection{BKT transition} 

\begin{figure}[b]
\centerline{\includegraphics[width=8.5cm]{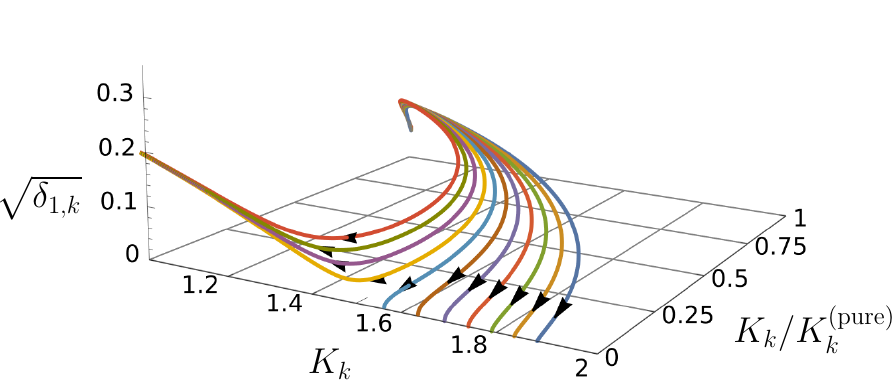}}
\caption{Flow trajectories projected onto the three-dimensional space $(K_k,K_k/\Kkpure,\sqrt{\delta_{1,k}})$, obtained for $\KLL=1$ and $\alpha=2$.} 
\label{fig_flowK1} 
\end{figure}

\begin{figure}[b]
\centerline{\includegraphics[width=7.25cm]{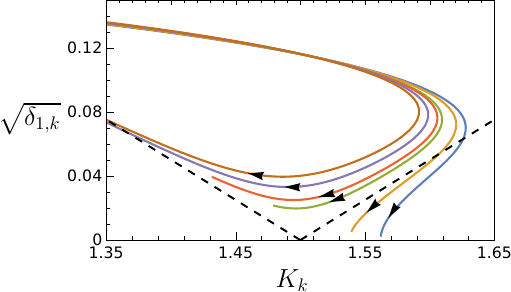}}
\caption{Flow trajectories passing close to the BKT point obtained for $\KLL=1$ and various values of the disorder strength. The flow diagram is projected onto the plane $(K_k,\sqrt{\delta_{1,k}})$ and the dashed lines show the two separatrix of the BKT flow~(\ref{bktflow}).}
\label{fig_pflowK1} 
\end{figure}

Figure~\ref{fig_flowK1} shows some RG trajectories in the three-dimensional space defined by $K_k$, $K_k/\Kkpure$ and $\sqrt{\delta_{1,k}}$ ($\alpha=2$). The initial condition corresponds to $\KLL=1$ and various values of disorder near the critical value so that all trajectories pass close to the BKT point defined by $K=3/2$, $K/K^{\rm (pure)}=0$ and $\delta(u)=0$. The ratio $K_k/\Kkpure$ rapidly decreases and long-range hopping plays no role at low energy since $K_k/\Kkpure\to 0$ for $k\to 0$; the nearly critical trajectories are ultimately located in the plane $(K_k,\sqrt{\delta_{1,k}})$. 

Figure~\ref{fig_pflowK1} shows the trajectories projected onto the plane $(K_k,\sqrt{\delta_{1,k}})$. The initial increase of $K_k$ is due to the effect of long-range hopping and occurs in a regime where $K_k/\Kkpure$ is still of order unity (Fig.~\ref{fig_flowK1}). To analyze the flow near the BKT point, one can retain only the first harmonic $\delta_{1,k}$, which is the only one being relevant when $K_k$ is near $3/2$. The flow equations then read 
\beq
\begin{split} 
\dt \delta_{1,k} &= (-3 + 4K_k l_1) \delta_{1,k} + 4\pi\bar l_2 \delta_{1,k}^2 , \\ 
\dt K_k &= - 2\pi \bar m_\tau \delta_{1,k} K_k . 
\end{split}
\eeq 
The relevant variables for the study of the nearly critical trajectories are $x=K_k-3/2$ and $y=\sqrt{\delta_{1,k}}$. To obtain the flow equations to second order in these variables, it is sufficient to evaluate the threshold functions $l_1$ and $\bar m_\tau$ for $\theta_k=\dt\ln K_k=0$, which gives the universal value $l_1=1/2$ (independent of the choice of the function $r$ in~(\ref{Rdef})). This leads to the standard equations of the BKT transition~\cite{Kosterlitz74,Chaikin_book}, 
\beq 
\begin{split} 
\dt y &= xy , \\
\dt x &= -3\pi \bar m_\tau y^2 ,
\end{split}
\label{bktflow} 
\eeq 
where $\bar m_\tau\equiv \bar m_\tau(\theta_k=0)<0$ is a nonuniversal number that depends on the choice of the cutoff function $R_k$. The critical trajectory corresponds to $y=\pm x/\sqrt{3\pi|\bar m_\tau|}$, in very good agreement with the numerical solution of the full flow equations (Fig.~\ref{fig_pflowK1}).

\subsubsection{Flow diagram} 
\label{subsubsection_flow_diagram} 

\begin{figure}
\centerline{\includegraphics[width=9cm]{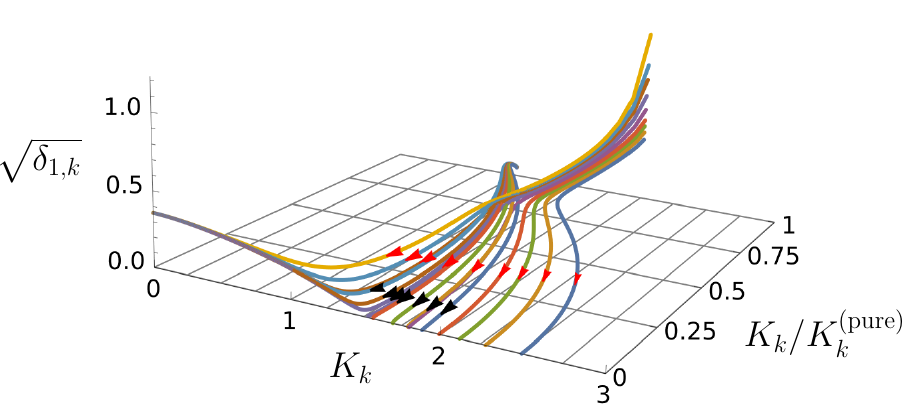}}
\caption{Flow trajectories in the three-dimensional space $(K_k,K_k/\Kkpure,\sqrt{\delta_{1,k}})$, obtained for $\KLL=1$ and $\KLL=2$ with $\alpha=2$. The black (dark) and red (light) arrows correspond to flow trajectories obtained for $\KLL=1$ and $\KLL=2$, respectively. The attractive fixed point located at $(K=0,K/K^{\rm (pure)}=0,\sqrt{\delta^*_1})$ describes the Bose-glass phase.}
\label{fig_flowK1K2} 
\end{figure}

In Fig.~\ref{fig_flowK1K2}, we show several flow trajectories in the three-dimensional space $(K_k,K_k/\Kkpure,\sqrt{\delta_{1,k}})$, corresponding to $\KLL=1$ and $\KLL=2$ and various values of the disorder intensity ($\alpha=2$). In both cases, the trajectories flow to the superfluid phase if the disorder is sufficiently weak, but end up in the Bose-glass phase for strong enough disorder. For $\KLL=2$, a very strong value of disorder is necessary to destabilize the superfluid phase. Long-range hopping plays no role, i.e. $K_k/\Kkpure\simeq 0$, near the BKT point and up to the Bose-glass fixed point. The latter is characterized by
a finite compressibility $\kappa=1/\pi^2Z_x=\KLL/\pi\vLL$, a vanishing Luttinger parameter $K=0$, associated with pinning of the $\varphi$ field, and a singular disorder correlator $\delta^*(\phi_a-\phi_b)$ whose functional dependence assumes a cuspy form related to the existence of metastable states. At nonzero momentum scale $k$, quantum tunneling between the ground state and low-lying metastable states leads to a rounding of the cusp singularity into a quantum boundary layer. The latter controls the low-energy dynamics and yields a (dissipative) conductivity vanishing as $\w^2$ in the low-frequency limit. We refer to Refs.~\cite{Dupuis19,Dupuis20} for a more detailed discussion. 

\begin{figure}[b]
	\centerline{\includegraphics[width=8cm]{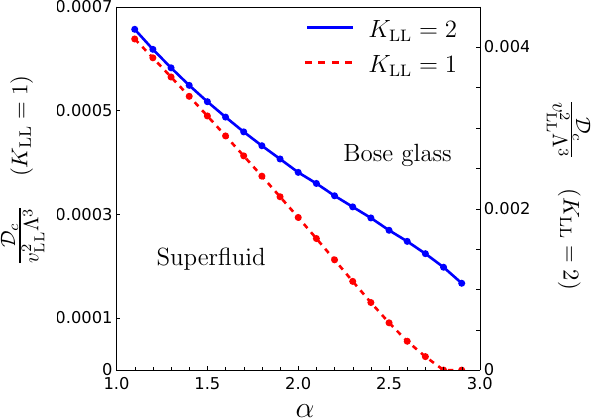}}
	\caption{Critical value $\calD_c$ of the disorder strength separating the Bose-glass phase ($\calD>\calD_c$) from the superfluid phase ($\calD<\calD_c$) as a function of $\alpha$, for $\KLL=2$ and $\KLL=1$.}
	\label{fig_phasedia} 
\end{figure}

For given values of the parameters of the microscopic action~(\ref{action5}), it is possible to find the critical value $\calD_c$ of the disorder strength separating the superfluid phase from the Bose-glass phase, as a function of $\alpha$. The phase diagram is shown in Fig.~\ref{fig_phasedia} for $\KLL=1$ and $\KLL=2$. In both cases, the transition line $\calD_c(\alpha)$ is a monotonously decreasing function of $\alpha$, but the order of magnitude of $\calD_c$ strongly depend on the value $\KLL$. As pointed out earlier, the fact that we consider $\KLL$ and $\alpha$ as independent parameters does not affect the nature of the superfluid--Bose-glass transition but may change the phase diagram. The monotonously decreasing transition line $\calD_c(\alpha)$ is qualitatively correct only if $\KLL(\alpha)$ weakly depends on $\alpha$. In the opposite case, there is no guarantee that our result provides a reliable (qualitative) estimate of the transition line of the original lattice model. Interestingly however, the transition line shown in Fig.~\ref{fig_phasedia} is in qualitative agreement with the result of numerical simulations of Ref.~\cite{Gupta23} even though the nature of the transition, as already stressed, is found to be different in the two approaches.

\section{conclusion}

A boson system with long-range hopping is characterized by an enhanced superfluid character, which manifests itself by a spontaneous breaking of the U(1) symmetry and a superfluid stiffness diverging in the small-momentum limit. Consequently, a minimum disorder strength is required to destabilize the superfluid phase in favor of a Bose-glass phase. Nevertheless, this is the only property of the system that differentiates it from a boson system with short-range hopping. Long-range hopping is indeed not resistant to disorder and plays no role in the low-energy limit of the FRG flow equations. This implies that the superfluid phase, when stable, is characterized by a sound mode with linear dispersion $\w\sim |q|$ for $q\to 0$ and the superfluid--Bose-glass transition remains in the BKT universality class.

Except for the stability of the superfluid phase at weak disorder, our conclusions disagree with a recent quantum Monte Carlo study of a lattice hard-core boson model with long-range hopping~\cite{Gupta23}. Nevertheless, we do not expect any fundamental difference between hard-core ($U\to\infty$) and soft-core ($U<\infty$) bosons since in both cases the system behaves as a Luttinger liquid in the absence of disorder and long-range hopping; only the value of the Luttinger parameter differs, with $\KLL=1$ for hard-core bosons. The numerical study of Ref.~\cite{Gupta23} seems to indicate that the pure system behaves as a regular Luttinger liquid with no spontaneous breaking of the U(1) symmetry and power-law decaying superfluid correlations (characterized by an effective Luttinger parameter), which contradicts results obtained from bosonization, the density-matrix renormalization group and spin-wave analysis~\cite{Maghrebi17,Frerot17} in the equivalent spin-$\half$ XXZ model. In addition, the authors of Ref.~\cite{Gupta23} find that the superfluid--Bose-glass transition is not in the BKT universality class but rather corresponds to a usual second-order phase transition with a power-law diverging correlation length. This conclusion contradicts the analysis reported in the present manuscript, and also seems incompatible with the authors' own conclusion that a boson system with long-range hopping would behave as a Luttinger liquid in the absence of disorder. The failure of the numerical analysis to detect the spontaneous U(1) symmetry breaking of the pure system could be due to the crossover length $1/q_x$, above which long-range hopping dominates, being larger than the system size, but this would not explain why the superfluid--Bose-glass transition is not found in the BKT universality class if the pure system appears to behave as a regular Luttinger liquid.

\section*{Acknowledgment}
I thank G. Pupillo for discussions on the work reported in Ref.~\cite{Gupta23}. 

\appendix
 
\section{Correlation functions of the pure system} 
\label{app} 

Correlation functions of the pure system can be obtained from the propagator of the $\varphi$ and $\theta$ fields derived from the action~(\ref{action3}), 
\beq 
\begin{split}
	G_{\varphi\varphi}(Q) &= \frac{\pi v(q) K(q)}{\w^2+v(q)^2 q^2}, \\ 
 G_{\theta\theta}(Q) &= \frac{\pi v(q)/ K(q)}{\w^2+v(q)^2 q^2} . 
\end{split}
\eeq
The long-wavelength part of the density-density correlation function is given by 
\beq 
\chi_{\rho\rho}(Q) = \frac{q^2}{\pi^2} G_{\varphi\varphi}(Q) = \frac{q^2}{\pi} \frac{v(q) K(q)}{\w^2+v(q)^2 q^2}  ,
\eeq 
which gives the compressibility 
\beq 
\kappa = \lim_{q\to 0} \chi_{\rho\rho}(q,i\w=0) = \lim_{q\to 0}  \frac{K(q)}{\pi v(q)} . 
\eeq 
The conductivity 
\beq 
\sig(q,\w)=\frac{\w+i0^+}{iq^2}\chi_{\rho\rho}(q,\w+i0^+)
\eeq 
can be obtained from $\chi_{\rho\rho}$ using gauge invariance~\cite{NDbook1}, and its real part is given by~(\ref{Resig}). 
 
Phase fluctuations are bounded, 
\begin{align}
\mean{\theta(x,\tau)^2} &= \int_{q,\w}  G_{\theta\theta}(Q) 
= \frac{\pi \vLL}{2\KLL} \int_q \frac{1}{v(q)|q|} \nonumber\\ 
& \sim \int_0 dq \frac{1}{|q|^{\frac{\alpha-1}{2}}} 
\end{align}
(we use the notation $\int_\w=\int \frac{d\w}{2\pi}$ and $\int_q=\int \frac{dq}{2\pi}$),
since the momentum integral is infrared convergent when $\alpha<3$ (a UV momentum cut-off is implicit). This implies the spontaneous breaking of the U(1) invariance [Eq.~(\ref{U1broken})]. The long-distance behavior of the connected superfluid correlation function [Eq.~(\ref{GSFc})] 
\beq 
G(x) = e^{G_{\theta\theta}(x,0)-G_{\theta\theta}(0,0)} - e^{-G_{\theta\theta}(0,0)} 
\eeq 
can be computed using 
\begin{align}
G_{\theta\theta}(x,0)-G_{\theta\theta}(0,0) &= \int_{q,\w} [\cos(qx)-1] G_{\theta\theta}(Q) \nonumber\\ 
&= \frac{\pi \vLL}{2\KLL} \int_q \frac{\cos(qx)-1}{v(q)|q|} \nonumber\\ 
&\simeq - C|x|^{\frac{\alpha-3}{2}} + C' ,
\end{align}
where $C$ and $C'$ are constants (with $C>0$). 

Finally, using 
\begin{align}
	G_{\varphi\varphi}(x,0)-G_{\varphi\varphi}(0,0) &= \int_{q,\w} [\cos(qx)-1] G_{\varphi\varphi}(Q) \nonumber\\ 
	&= \frac{\pi }{2} \int_q \frac{K(q)}{|q|} [\cos(qx)-1] \nonumber\\ 
	&\simeq -C|x|^{\frac{3-\alpha}{2}}  ,
\end{align} 
we obtain the long-distance behavior of the $q=2\pi\rho_0$ density-density correlation function 
\beq 
\chi(x) = e^{4G_{\varphi\varphi}(x,0)-4G_{\varphi\varphi}(0,0)} 
\eeq 
given in Eq.~(\ref{chi}).

\section{Threshold functions} 
\label{app_threshold} 

The threshold functions involved in the RG equation of the two-replica potential are defined by 
\beq 
\begin{split}
l_1 &= \pi \int_{\tilde Q} \tilde P_k(\tilde Q)^2 \dt \tilde R_k(\tilde Q) , \\ 
\bar l_2 &= 2\pi \int_{\tilde q} \tilde P_k(\tilde q,0)^3 \dt \tilde R_k(\tilde q,0) ,
\end{split}
\eeq 
where 
\begin{align} 
\tilde P_k(\tilde Q) ={}& \biggl\{ \tilde q^2[1+r(\tilde Q^2)] 
+ \tw^2 \left[  \frac{v_k^2}{v_k(q)^2} + r(\tilde Q^2)\right] \biggr\}^{-1} 
\end{align} 
is the dimensionless propagator obtained from $\Gamma_{1,k}[\phi_a]$, with $\tilde Q=(\tilde q,i\tw)$, $\tilde Q^2=\tilde q^2+\tw^2$, $\tilde q=q/k$ and $\tw=\w/v_kk$. The dimensionless cutoff function and its time derivative are given by 
\beq 
\begin{split}
\tilde R_k(\tilde Q) ={}& \tilde Q^2 r(\tilde Q^2) , \\ 
\dt \tilde R_k(\tilde Q) ={}& - 2 \bigl\{ [ r(\tilde Q^2) + \tilde Q^2 r'(\tilde Q^2) ] (z_k-1) \tw^2 \\ & + \tilde Q^4 r'(\tilde Q^2) \bigr\} . 
\end{split}
\eeq 
The threshold function $l_1$ is not defined for $Y_k=0$, i.e. when $k=\Lambda$, since the integral over $\tilde Q$ is infrared divergent. A nonzero $Y_k$ appearing for $k<\Lambda$, we assume a very small initial value $Y_\Lambda\sim 10^{-3}$, which makes $l_1$ finite for $k=\Lambda$. The flow is essentially independent of the chosen value for $Y_\Lambda$.   

The threshold function $\bar m_\tau$ appearing in the expression~(\ref{thetak}) of $\theta_k$ is defined by 
\beq 
\begin{split}
\bar m_\tau &= \partial_{\tw^2} \bar l_1(i\tw)\bigl|_{\tw=0} , \\ 
\bar l_1(i\tw) &= \pi \int_{\tilde q} \tilde P_k(\tilde Q)^2 \dt \tilde R_k(\tilde Q) .
\end{split}
\eeq 
It can be written in the form 
\begin{align} 
\bar m_\tau ={}& \pi \int_{\tilde q} \bigl\{ 2 \tilde P_k(\tilde Q) [ \partial_{\tw^2}\tilde P_k(\tilde Q) ] \dt\tilde R_k(\tilde Q) \nonumber\\ & + \tilde P_k(\tilde Q)^2  \partial_{\tw^2} \dt\tilde R_k(\tilde Q) \bigr\}_{\tw=0} , 
\end{align} 
where 
\begin{align}
&\partial_{\tw^2}\tilde P_k(\tilde Q) = - \left[ \frac{v_k^2}{v_k(q)^2} + r(\tilde Q^2) + \tilde Q^2 r'(\tilde Q^2) \right] \tilde P_k(\tilde Q)^2 , \nonumber\\ 
&\partial_{\tw^2} \dt\tilde R_k(\tilde Q) \bigr|_{\tw=0} = -2 \bigl\{  (z_k-1) [r(\tilde q^2) + \tilde q^2 r'(\tilde q^2) ] \nonumber\\ & \hspace{3cm} + \tilde q^2 [ 2 r'(\tilde q^2) + \tilde q^2 r''(\tilde q^2) ] \bigr\} .
\end{align}

%


\end{document}